\documentclass[11pt]{article}
\usepackage{amssymb,cite}
\makeatletter
\@addtoreset{equation}{section}
\makeatother

\def\ben{\begin{equation}}
\def\een{\end{equation}}

   \let\i=\iota

 \def\bd{\begin{document}} \def\ed{\end{document}}
\def\ds{\documentstyle} \let\fr=\frac \let\bl=\bigl \let\br=\bigr
\let\Br=\Bigr \let\Bl=\Bigl
\let\bm=\bibitem
\let\na=\nabla
\let\pa=\partial \let\ov=\overline
\newcommand{\be}{\begin{equation}}
\newcommand{\ee}{\end{equation}}
\def\ba{\begin{array}}
\def\ea{\end{array}}
\def\ft#1#2{{\textstyle{{\scriptstyle #1}\over {\scriptstyle #2}}}}
\def\fft#1#2{{#1 \over #2}}
\def\del{\partial}
\def\vp{\varphi}
\def\sst#1{{\scriptscriptstyle #1}}
\def\oneone{\rlap 1\mkern4mu{\rm l}}
\def\td{\tilde}
\def\wtd{\widetilde}
\def\ie{\rm i.e.\ }
\def\dalemb#1#2{{\vbox{\hrule height .#2pt
        \hbox{\vrule width.#2pt height#1pt \kern#1pt
                \vrule width.#2pt}
        \hrule height.#2pt}}}
\def\square{\mathord{\dalemb{6.8}{7}\hbox{\hskip1pt}}}
\newcommand{\ho}[1]{$\, ^{#1}$}
\newcommand{\hoch}[1]{$\, ^{#1}$}
\newcommand{\bea}{\begin{eqnarray}}
\newcommand{\eea}{\end{eqnarray}}

\title{A Note on Particles and Scalar Fields in Nutty Spacetimes}

\newcommand{\auth}{Muraari Vasudevan}

\begin{document}

\begin{flushright}

Alberta Thy 13-05\\
PACS numbers: 04.50.+h, 98.80.Cq\hfill\\
November\  2005
\end{flushright}

\vspace{10pt}

\begin{center}
{\large {\bf A Note on Particles and Scalar Fields in Higher
Dimensional Nutty Spacetimes
            }}

\vspace{20pt}
\auth

\vspace{10pt}

{\it  Theoretical Physics Institute, University of Alberta,\\
Edmonton, Alberta  T6G 2J1, Canada}

{\it
    {\rm E-mail: \texttt{mvasudev@phys.ualberta.ca}}
  }

%\vspace{10pt} {\hoch{\dagger}\brussels}

\vspace{40pt}

\underline{ABSTRACT}
\end{center}

In this note, we study the integrability of geodesic flow in the
background of a very general class of spacetimes with NUT-charge(s)
in higher dimensions. This broad set encompasses multiply
NUT-charged solutions, electrically and magnetically charged
solutions, solutions with a cosmological constant, and time
dependant bubble-like solutions. We also derive first-order
equations of motion for particles in these backgrounds. Separability
turns out to be possible due to the existence of non-trivial
irreducible Killing tensors. Finally, we also examine the
Klein-Gordon equation for a scalar field in these spacetimes and
demonstrate complete separability.

\pagebreak

\section{Introduction}
Taub-NUT solutions arise in a very wide variety of situations in
both string theory and general relativity. NUT-charged spacetimes,
in general, are studied for their unusual properties which typically
provide rather unique counterexamples to many notions in Einstein
gravity. They are also widely studied in the context of issues of
chronology protection in the AdS/CFT correspondence. Understanding
the nature of geodesics in these backgrounds, as well as scalar
field propagation, could prove to be very interesting in further
exploration of these spacetimes.

There is a strong need to understand explicitly the structure of
geodesics in the background of black holes in Anti-de Sitter space
in the context of string theory and the AdS/CFT correspondence. This
is due to the recent work in exploring black hole singularity
structure using geodesics and correlators in the dual CFT on the
boundary \cite{Shen1, Shen2, Shen3, Shen4, Shen5, Shen6}. Black
holes with charge are particularly interesting for this type of
analysis since the charges are reinterpreted as the R-charges of the
dual theory. The class of solutions dealt with in this paper also
include black holes that carry both NUT and electric charges in
various dimensions, and could prove very interesting in this sort of
analysis.

In this paper we explore a very general metric describing a wide
variety of spacetimes with NUT charge(s). In addition further
metrics can also be obtained from these through various analytic
continuations (which does not affect separability as demonstrated
for these class of metrics). As such, the study of separability in
this set of spacetimes encompasses the cases of both singly and
multiply NUT-charged solutions, electrically and magnetically
charged solutions with NUT parameter(s), solutions with a
cosmological constant and NUT parameters(s), and time dependant
bubble-like NUT-charged solutions. Many of these describe very
interesting gravitational instantons. Some of these solutions
include static backgrounds, while others are time-dependant and
provide very interesting backgrounds for studying both string theory
and general relativity. Some of these solutions, especially the
bubble-like ones, are particularly interesting in the context of
string theory as they arise in the context of topology changing
processes. e.g. they show up as possible end states for Hawking
evaporation., and they show up in transitions of black strings in
closed string tachyon condensation.

We study the separability of the Hamilton-Jacobi equation in these
spacetimes, which can be used to describe the motion of classical
massive and massless particles (including photons). We use this
explicit separation to obtain first-order equations of motion for
both massive and massless particles in these backgrounds. The
equations are obtained in a form  that could be used for numerical
study, and also in the study of black hole singularity structure
using geodesic probes and the AdS/CFT correspondence. We also study
the Klein-Gordon equation describing the propagation of a massive
scalar field in these spacetimes. Separation again turns out to be
possible with the usual multiplicative ansatz.

Separation is possible for both equations in these metrics due to
the existence of non-trivial second-order Killing tensors. The
Killing tensors, in each case, provides an additional integral of
motion necessary for complete integrability.

There has been a lot of work recently dealing with geodesics and
integrability in black hole backgrounds in higher dimensions both
with and without the presence of a cosmological constant
\cite{frolov1, frolov2, VSP1, KL, VSP2, MV, VS1, ZGLP, MDS1, KL2}.
Of particular note in the context of this paper are \cite{MV,ZGLP}
which deal with black holes with NUT parameters in some special
cases. This work extends, and generalizes, some of the results
obtained in these papers.

\section{Overview of the Metrics}

The class of metrics dealt with in this paper, and their
generalizations obtained via analytic continuations, have been
constructed and analyzed in \cite{Nut1, Nut2, Nut3, Nut4, Nut5,
Nut6}, as well as some references contained therein. We will very
briefly describe the metrics, and some of the various types of
spacetimes that can be obtained from them. As mentioned earlier,
separability for all the metrics is addressed by dealing with the
class we do here, since analytic continuations do not affect
separability of either the Hamilton-Jacobi or Klein-Gordon equation
(though they do affect the physical interpretations of the various
variables and their associated conserved quantities).

The general spacetimes we study are described by the metrics
\begin{eqnarray}
ds^2=-F(r)\left[dt+\sum_{i=1}^p 2N_i
f_i(\theta_i)d\phi_i\right]^2+\frac{dr^2}{F(r)}+\sum_{i=1}^p(r^2+N_i^2)(d\theta_i^2+g_i^2(\theta_i)d\phi_i^2)
\, . \label{metrics}
\end{eqnarray}

A very general class of metrics in even dimensions where the
$(\phi_i,\theta_j)$ sector has the form $M_1\times M_2\times\ldots
\times M_p$, with each $M_i$ a two dimensional space of constant
curvature $\delta_i$. In this case the functions are given by
\begin{eqnarray}
&&\delta_i=1:\quad f_i(\theta_i)=-\cos \theta_i \, , \quad
g_i^2(\theta_i)=\sin^2\theta_i \, , \nonumber \\
&&\delta_i=0:\quad f_i(\theta_i)=-\theta_i \, , \quad
g_i^2(\theta_i)=1 \, , \nonumber \\
&&\delta_i=-1:\quad f_i(\theta_i)=-\cosh \theta_i \, , \quad
g_i^2(\theta_i)=\sinh^2\theta_i \, , \label{funcs}
\end{eqnarray}
and an expression for $F(r)$ can be found in \cite{Nut5} along with
a detailed description. Generalizations to include electric charge
are obtained by suitably modifying $F(r)$, and can be found in
\cite{Nut4, Nut6}. Metrics describing ``bubbles of nothing" also
fall under this class and can be found in \cite{Nut3}. Examples of
NUT-charged spacetimes in cosmological backgrounds also fall in this
framework and can be found in \cite{Nut3}.

For the purposes of analyzing separability, some odd dimensional
NUT-charged spacetimes also fall under this category. For instance
in five dimensions (i.e $p=2$) a NUT charged spacetime is obtained
by taking $g_2(\theta_2)=0$ and $N_2=0$, i.e. a metric of the form
\begin{eqnarray}
ds^2=-F(r)(dt-2N_1\cosh\theta_1 d\phi_1)^2 +\frac{dr^2}{F(r)} +
(r^2+N_1^2)(d\theta_1^2 + \sinh^2\theta_1 d\phi_1^2) + r^2
d\theta_2^2 \, .
\end{eqnarray}
This describes a spacetime in an AdS background; similar dS and flat
background spacetimes can be obtained by following the prescriptions
in (\ref{funcs}) while maintaining $g_2(\theta_2)=0$ and $N_2=0$.
Generalizations to higher odd dimensional spacetimes are obvious.

Various twists of these spacetimes can also be obtained through
analytic continuations. For instance, using the prescriptions
$t\rightarrow i\theta, \theta\rightarrow it$, we can obtain
time-dependant bubbles. In five dimensions in an AdS background,
some examples obtained via this prescription, and a few other
suitable obvious variable redefinitions are
\begin{eqnarray}
ds^2&=& F(r)(d\theta_1 + 2N_1 \cos t d\phi)^2 + \frac{dr^2}{F(r)} +
(r^2+N_1^2)(-dt^2 +\sin^2t d\phi^2)+r^2d\theta_2^2 \, , \nonumber \\
ds^2&=& F(r)(d\theta_1 + 2N_1 \sinh \phi dt)^2 + \frac{dr^2}{F(r)} +
(r^2+N_1^2)(d\phi^2 -\cosh^2\phi dt^2)+r^2d\theta_2^2 \, ,
\nonumber\\
 ds^2&=& F(r)(d\theta_1 + 2N_1 \cosh \phi dt)^2 +
\frac{dr^2}{F(r)} + (r^2+N_1^2)(d\phi^2 -\sinh^2\phi
dt^2)+r^2d\theta_2^2 \, , \nonumber
\\
 ds^2&=& F(r)(d\theta_1 + 2N_1 e^\phi dt)^2 + \frac{dr^2}{F(r)} +
(r^2+N_1^2)(d\phi^2 -e^{2\phi} dt^2)+r^2d\theta_2^2 \, .
\end{eqnarray}

For future use, we give the determinant of the metric
(\ref{metrics})
\begin{eqnarray}
g=-\prod_{i=1}^p (r^2+N_i^2)^2g_i^2(\theta_i) \, . \label{detnut}
\end{eqnarray}
The components of the inverse metric are
\begin{eqnarray}
g^{tt}&=&\sum_{i=1}^p \frac{4N_i^2
f_i^2(\theta_i)}{(r^2+N_i^2)g_i^2(\theta_i)} -\frac{1}{F(r)} \, ,
\nonumber \\
g^{t\phi_i}&=& -\frac{2N_if_i(\theta_i)}{g_i^2(\theta_i)(r^2+N_i^2)}
\, , \nonumber \\
g^{\phi_i \phi_j} &=&\frac{\delta_{ij}}{(r^2+N_i^2)g_i^2(\theta_i)}
\, , \nonumber \\
g^{rr}&=&F(r) \, , \nonumber \\
g^{\theta_i\theta_j} &=& \frac{\delta_{ij}}{r^2+N_i^2} \, .
\label{invnut}
\end{eqnarray}
These formulae are somewhat tedious to derive, but can be proved
using a few Maple calculations, and then using mathematical
induction \cite{Map}.

\section{The Hamilton-Jacobi Equation and Separability}

The Hamilton-Jacobi equation in a curved background is given by \be
-\frac{\partial S}{\partial \lambda} = H = \frac{1}{2} g^{\mu \nu }
\frac{\partial S}{\partial x^{\mu}} \frac{\partial S}{\partial
x^{\nu}} \,, \label{HJ} \ee where $S$ is the action associated with
the particle and $\lambda$ is some affine parameter along the
worldline of the particle. Note that this treatment also
accommodates the case of massless particles, where the trajectory
cannot be parametrized by proper time.

\subsection{Separability}
We can attempt a separation of coordinates as follows. Let
\begin{equation}
S=\frac{1}{2}m^2 \lambda -Et + \sum_{i=1}^p L_{\phi_i} \phi_i +
\sum_{i=1}^p S_{\theta_i} (\theta_i) + S_r (r)\,. \label{ansatznut}
\end{equation}
$t$ and the $\phi_i$ are cyclic coordinates, so their conjugate
momenta are conserved. The conserved quantity associated with time
translation is the energy $E$, and those with rotation in the
$\phi_i$ are the corresponding angular momenta $L_{\phi_i}$. Then,
using the components of the inverse metric (\ref{invnut}), the
Hamilton-Jacobi equation (\ref{HJ}) is written to be
\begin{eqnarray}
-m^2&=& \sum_{i=1}^p \frac{4N_i^2 f_i ^2
(\theta_i)}{(r^2+N_i^2)g_i^2(\theta_i)}E^2-\frac{E^2}{F(r)}-\sum_{i=1}^p
\frac{4N_i
f_i(\theta_i)}{(r^2+N_i^2)g_i^2(\theta_i)}(L_{\phi_i})(-E) \nonumber
\\
&&+\sum_{i=1}^p \frac{1}{(r^2+N_i^2)g_i^2(\theta_i)}L^2_{\phi_i} +
F(r)\left[\frac{dS_r(r)}{dr}\right]^2 + \sum_{i=1}^p \frac{1}{r^2+
N_i^2} \left[\frac{dS_{\theta_i}(\theta_i)}{d\theta_i}\right]^2 \, .
\end{eqnarray}

After some manipulation, we can recursively separate out the
equation into
\begin{eqnarray}
-m^2&=& -\frac{E^2}{F(r)}+F(r)\left[\frac{dS_r(r)}{dr}\right]^2 +
\sum_{i=1}^p \frac{K_i}{r^2+N_i^2} \, , \nonumber \\
K_i&=&\left[\frac{dS_{\theta_i}(\theta_i)}{d\theta_i}\right]^2 +
\left[\frac{L_{\phi_i}+2N_i f_i(\theta_i)E}{g_i(\theta_i)}\right]^2
\, . \label{sepnut}
\end{eqnarray}

For future reference we will use the notation $K=\sum_{i=1}^p K_i$.
Also note that for the metrics obtained through analytic
continuations discussed earlier, the issue of separability is
clearly not affected. However, for an analytic continuation of the
form $t\rightarrow i \theta, \theta\rightarrow it$, we need to
replace $E\rightarrow -iL_\theta$, and the energy is no longer
conserved as we have a time dependant background. However, now the
angular momentum $L_\theta$ associated to $\theta$ is conserved.
Similar substitutions need to be made for any other analytic
continuations or variable redefinitions used to define the new
metrics.

\subsection{The Equations of Motion}

To derive the equations of motion, we will write the separated
action $S$ from the Hamilton-Jacobi equation in the following form:
\bea S&=&\frac{1}{2}m^2 \lambda -Et +\sum_{\i=1}^p L_{\phi_i} \phi_i
+ \int ^r \sqrt{\mathcal{R}(r')} dr' + \sum_{i=1}^p \int ^
{\theta_i} \sqrt{\Theta_i(\theta'_i)}d\theta'_i \, , \eea where
\begin{eqnarray}
F(r)\mathcal{R}(r)&=&-\sum_{i=1}^p \frac{K_i}{r^2+N_i^2}+\frac{E^2}{F(r)} - m^2 \, , \nonumber\\
\Theta_i(\theta_i)&=&K_i-\left[\frac{L_{\phi_i}+2N_i
f_i(\theta_i)E}{g_i(\theta_i)}\right]^2\,.
\end{eqnarray}

To obtain the equations of motion, we differentiate $S$ with respect
to the parameters $m^2,K_i,E,L_{\phi_i}$ and set these derivatives
to equal other constants of motion. However, we can set all these
new constants of motion to zero (following from freedom in choice of
origin for the corresponding coordinates, or alternatively by
changing the constants of integration). Following this procedure, we
get the following equations of motion:
\begin{eqnarray}
\frac{\pa S}{\pa m^2}&=&0 \Rightarrow \lambda = \int \frac{dr}{F(r)\sqrt{\mathcal R(r)}} \, , \nonumber \\
\frac{\pa S}{\pa K_i}&=&0 \Rightarrow \int \frac{d\theta_i}{\sqrt{\Theta_i}} = \int \frac{1}{(r^2+N_i^2)}\frac{dr}{F(r)\sqrt{\mathcal R(r)}} \, , \nonumber \\
\frac{\pa S}{\pa L_{\phi_i}}&=&0 \Rightarrow \phi_i=\int
\frac{L_{\phi_i}+2N_if_i(\theta_i)E}{g_i^2(\theta_i)}
\frac{d\theta_i}{\sqrt{\Theta_i(\theta_i)}} \, , \label{inteqs} \\
\frac{\pa S}{\pa E}&=&0 \Rightarrow
t=\int\frac{E}{F^2(r)}\frac{dr}{\sqrt{\mathcal{R}(r)}}
-\sum_{i=1}^{p}\int\frac{2N_iL_{\phi_i}f_i(\theta_i)+4N_i^2f_i^2(\theta_i)E}{g_i^2(\theta_i)}
\frac{d\theta_i}{\sqrt{\Theta_i(\theta_i)}} \,. \nonumber
\end{eqnarray}
It is often more convenient to rewrite these in the form of
first-order differential equations obtained from (\ref{inteqs}) by
direct differentiation with respect to the affine parameter:
\begin{eqnarray}
\frac{dr}{d\lambda} &=& F(r) \sqrt{\mathcal R (r)} \, , \nonumber \\
\frac{d \theta_i}{d\lambda} &=& \frac{\sqrt{\Theta_i(\theta_i)}}{r^2+N_i^2} \, , \nonumber \\
\frac{d\phi_i}{d\lambda}&=&\frac{L_{\phi_i}+2N_i
f_i(\theta_i)E}{g_i^2(\theta_i)(r^2+N_i^2)} \, ,\nonumber \\
\frac{dt}{d\lambda}&=& \frac{E}{F(r)}-\sum_{i=1}^p
\frac{2N_iL_{\phi_i}f_i(\theta_i)+4N_i^2f_i^2(\theta_i)E}{g_i^2(\theta_i)(r^2+N_i^2)}\,
. \label{eqns}
\end{eqnarray}

\section{Dynamical Symmetry}
The general class of metrics discussed here are stationary and
``axisymmetric"; i.e., $\partial / \partial t$ and the $\partial /
\partial \phi_i$  are Killing vectors and have associated
conserved quantities, $-E$ and $L_{\phi_i}$. In general, if $\xi$ is
a Killing vector, then $\xi ^{\mu} p_{\mu}$ is a conserved quantity,
where $p$ is the momentum of the particle. Note that this quantity
is first order in the momenta.

The additional constants of motion $K_i$ which allowed for complete
integrability of the equations of motion is not related to a Killing
vector from a cyclic coordinate. These constants are, rather,
derived from irreducible second-order Killing tensors in which
permit the complete separation of equations. Killing tensors are not
symmetries on configuration space, and cannot be derived from a
Noether procedure, and are rather, symmetries on phase space. They
obey a generalization of the Killing equation for Killing vectors
(which do generate symmetries in configuration space by the Noether
procedure) given by
\begin{eqnarray}
\mathcal K_{(\mu\nu;\rho)}=0 \, ,
\end{eqnarray}
where $\mathcal K$ is any second order Killing tensor, and the
parentheses indicate complete symmetrization of all indices.

The Killing tensors can be obtained from the expressions for the
separation constants $K_i$ in each case. If the particle has
momentum $p$, then the Killing tensor $\mathcal K_{\mu \nu}$ is
related to the constant $K$ via
\begin{eqnarray}
K=\mathcal K^{\mu \nu}p_{\mu}p_{\nu}=\mathcal K^{\mu
\nu}\frac{\partial S}{\partial x^\mu}\frac{\partial S}{\partial
x^\nu} \, . \label{kilten}
\end{eqnarray}
We can use the expression for the $K_i$ in terms of the the
$\theta_i$ equations.

For the Taub-NUT metrics analyzed above, the expression for $K_i$
from (\ref{sepnut}) is
\begin{eqnarray}
K_i&=&\left[\frac{dS_{\theta_i}(\theta_i)}{d\theta_i}\right]^2 +
\left[\frac{L_{\phi_i}+2N_i f_i(\theta_i)E}{g_i(\theta_i)}\right]^2
\, .
\end{eqnarray}
Thus, from (\ref{kilten}) we can easily read
\begin{eqnarray}
\mathcal K_i=\partial_{\theta_i}\otimes \partial_{\theta_i} +
\frac{1}{g_i(\theta_i)^2}\left[\partial_{\phi_i}\otimes
\partial_{\phi_i}+4N_i^2f_i^2(\theta_i)\partial_t\otimes\partial_t-2N_if_i(\theta_i)sym(\partial_{\phi_i}\otimes\partial_t)\right]
\end{eqnarray}
We can easily check using Maple\cite{Map}, that the Killings tensors
do satisfy the Killing equation.

Note that if any of the NUT parameters $N_k$ were zero, then the
corresponding Killing tensor $\mathcal K_k$ would simply be the
usual Killing tensor of the underlying two dimensional space $M_k$
(which is a reducible one in the case of a homogenous constant
curvature space $M_k$ as is the case for many situations here). In
general, however, a non-zero NUT parameter $N_k$ provides a
nontrivial coupling between the $(r,\phi_i,\theta_i)$ sectors, and
the existence of the Killing vectors $\partial_{\phi_i}$ and
$\partial_t$ along is not enough to ensure complete separability. It
is the existence of these nontrivial irreducible Killing tensors
$\mathcal K_i$ that provides the addition separation constants $K_i$
necessary for complete separation of each space $M_i$ from another
space $M_j$, as well as separation of the angular sectors completely
from the radial sector. These tensors are irreducible since they are
not simply linear combinations of tensor products of Killing vectors
of the spacetime.

\section{The Scalar Field Equation}
Consider a scalar field $\Psi$ in a gravitational background with the action
\begin{equation}
S[\Psi]=-\frac{1}{2}\int d^Dx\sqrt{-g}((\nabla \Psi)^2+ \alpha R \Psi ^2 + m^2
\Psi ^2 ) \,,
\end{equation}
where we have included a curvature dependent coupling. However, in
these (Anti)-de Sitter and flat backgrounds with charges, $R$ is
constant (proportional to the cosmological constant $\Lambda$). As a
result we can trade off the curvature coupling for a different mass
term. So it is sufficient to study the massive Klein-Gordon equation
in this background. We will simply set $\alpha=0$ in the following.
Variation of the action leads to the Klein-Gordon equation
\begin{equation}
\frac{1}{\sqrt{-g}}\partial _{\mu}(\sqrt{-g} g^{\mu \nu}\partial _{\nu} \Psi
)=m^2 \Psi \,.\label{KG1}
\end{equation}

Using the explicit expressions for the components of the inverse
metric (\ref{invnut}) and the determinant (\ref{detnut}), the
Klein-Gordon equation for a massive scalar field in this spacetime
can be written as
\begin{eqnarray}
m^2\Psi&=&\left[\sum_{i=1}^p \frac{4N_i^2
f_i^2(\theta_i)}{(r^2+N_i^2)g_i^2(\theta_i)} -\frac{1}{F(r)}\right]
\partial_t^2
\Psi-\sum_{i=1}^p\frac{4N_if_i(\theta_i)}{g_i^2(\theta_i)(r^2+N_i^2)}\partial_{t\phi_i}^2\Psi
\nonumber \\
&&+\sum_{i=1}^p
\frac{1}{(r^2+N_i^2)g_i^2(\theta_i)}\partial_{\phi_i}^2\Psi+\frac{1}{\prod_{i=1}^p(r^2+N_i^2)}\frac{\partial}{\partial
r}\left[\prod_{i=1}^p (r^2+N_i^2) F(r)\frac{\partial \Psi}{\partial
r}\right]\nonumber \\
&&+\sum_{i=1}^p
\frac{1}{(r^2+N_i^2)g_i(\theta_i)}\frac{\partial}{\partial \theta_i}
\left[g_i(\theta_i) \frac{\partial \Psi}{\partial \theta_i}\right]
\, .
\end{eqnarray}
We assume the usual multiplicative ansatz for the separation of the
Klein-Gordon equation
\begin{eqnarray}
\Psi=\Phi_r(r)e^{-iEt}e^{i\sum_{i=1}^p L_{\phi_i} \phi_i}
\prod_{i=1}^p \Phi_{\theta_i}(\theta_i) \, .
\end{eqnarray}
Then we can easily completely separate the Klein-Gordon equation as
\begin{eqnarray}
K_i&=&\frac{1}{g_i(\theta_i)\Phi_{\theta_i}(\theta_i)}\frac{d}{d\theta_i}\left[g_i(\theta_i)
\frac{d\Phi_{\theta_i}(\theta_i)}{d\theta_i}\right]-\left[\frac{L_{\phi_i}+2N_if_i(\theta_i)E}{g_i(\theta_i)}\right]^2
\, , \nonumber \\
-m^2&=&\frac{1}{\prod_{i=1}^p(r^2+N_i^2)}\frac{d}{d
r}\left[\prod_{i=1}^p (r^2+N_i^2) F(r)\frac{d \Phi_r(r)}{d
r}\right]+\frac{E^2}{F(r)}+\sum_{i=1}^p\frac{K_i}{r^2+N_i^2} \, .
\end{eqnarray}
where the $K_i$ are again separation constants. At this point we
have completely separated out the Klein-Gordon equation for a
massive scalar field in these spacetimes.

We note the role of the Killing tensors in the separation terms of
the Klein-Gordon equations in these spacetimes. In fact, the
complete integrability of geodesic flow of the metrics via the
Hamilton-Jacobi equation can be viewed as the classical limit of the
statement that the Klein-Gordon equation in these metrics also
completely separates.

\section*{Conclusions}
We studied the complete integrability properties of the
Hamilton-Jacobi and the Klein-Gordon equations in the background of
a very general class of Taub-NUT metrics in higher dimensions, which
include the cases of both singly and multiply NUT-charged solutions,
electrically and magnetically charged solutions with NUT
parameter(s), solutions with a cosmological constant and NUT
parameters(s), and time dependant bubble-like NUT-charged solutions,
and other very interesting gravitational instantons. Complete
separation of both the Hamilton-Jacobi and Klein-Gordon equations in
these backgrounds is demonstrated. This is due to the enlarged
dynamical symmetry of the spacetime. We construct the Killing
tensors in these spacetimes which explicitly permit complete
separation. We also derive first-order equations of motion for
classical particles in these backgrounds. It should be emphasized
that these complete integrability properties are a fairly
non-trivial consequence of the specific form of the metrics, and
generalize several such remarkable properties for other previously
known metrics.

Further work in this direction could include the study of
higher-spin field equations in these backgrounds, which is of
great interest, particularly in the context of string theory.
Explicit numerical study of the equations of motion for specific
values of the black hole parameters could lead to interesting
results. The geodesic equations presented can also readily be used
in the study of black hole singularity structure in an AdS
background using the AdS/CFT correspondence.

\end{document}